\documentclass[twocolumn,twocolappendix]{aastex63}
\usepackage{amsmath}	
\usepackage{amssymb}	
\usepackage{booktabs}
\usepackage[hyphenbreaks]{breakurl}
\usepackage{color}  
\usepackage{enumitem}
\usepackage{hyperref}
\usepackage{ifsym}
\usepackage{multirow,tabularx}
\usepackage{pifont}
\usepackage{relsize}
\usepackage{stix}
\usepackage{subfigure}
\usepackage{xcolor}
\usepackage{wasysym} 
\usepackage[hyphens]{url}

\setitemize{noitemsep,topsep=0pt,parsep=0pt,partopsep=0pt}
\setenumerate{noitemsep,topsep=0pt,parsep=0pt,partopsep=0pt,after=\vspace{\baselineskip}}
\received{November 26th, 2020}
\accepted{December 15th, 2020}
\submitjournal{BAAS}

\shorttitle{Personal, Scientific Website Creation}
\shortauthors{E. Moravec}

\begin{document}

\title{A Resource for Creating a Website to Promote Your Scientific Work}

\correspondingauthor{Emily Moravec}
\email{emily.moravec@asu.cas.cz}

\author[0000-0001-9793-5416]{Emily Moravec}
\affiliation{Astronomical Institute of the Czech Academy of Sciences, Bo\v cn\'i II 1401/1A, 14000 Praha 4, Czech Republic}

\begin{abstract}
Creating a website to promote one's scientific work has become commonplace in many scientific disciplines. A plethora of options exist for framework to generate your website content, hosting it, and registering a domain name. The goal of this document is to provide early career scientists (1) an overview of the current options for creating a website to promote their professional persona, and (2) general advice concerning website written content and one's web presence. To get a sense of how other scientists created their websites, I created a survey asking colleagues about the services they used to create their websites and advice they have for someone creating a website. I received 54 responses from 53 astronomers and one computer scientist of which 23 were in an academic position beyond postdoc (faculty, scientist, etc.), 1 was an individual research fellow, 4 were in their third postdoc, 4 were in their second postdoc, 16 were in their first postdoc, and 6 were graduate students. I report the results of this survey here.
\end{abstract}


\section{Introduction} \label{sect:intro}
Why as a scientist should you create a website to promote myself and my work? The simple answer is that other professionals can easily find you, learn about your work, and contact you. What comes up when someone searches your name + profession (or keyword about your work)? The following are a few exemplary scenarios in which another professional might Google your name: (1) potential employers, (2) collaborators and potential collaborators, (3) another professional trying to learn more about you and your work, (4) colleagues at a conference, (5) a colleague wanting to invite you to give a talk, (6) a journalist who may want to contact you, either about your work, or to get your comments for other work, (7) a non-scientist wanting to invite you for seminar/outreach, and (8) potential non-academic employers who want to get in touch. Creating a website about you and your work will allow the aforementioned colleagues to find your professional persona and content relevant to your work (as opposed to other content that may be less professional such as social media). But how does one create such a website?

Creating a website to promote one's scientific work has become commonplace in many scientific disciplines. Over the years, the number of options for generating content, hosting, and registering domain names for one's website has increased exponentially. However, if one does not have experience in web development, it can be overwhelming to digest the plethora of options, choose method, and create a website. The motivation for this document is providing early career scientists with a resource and reference to help get you started in creating your website. 

The three main decisions that you will need to make concerning your website are (1) the framework to generate your website content, (2) where to host your website, and (3) where to register your domain name. You need all three components to produce a website. 

First, you will need to decide the framework that you will use to generate your website content (not the written/creative content). Generally, the appearance and style of your website is dictated by a combination of HTML and CSS (see \href{https://www.turnwall.com/articles/html-and-css-work-together/}{this blog post} explaining HTML versus CSS). Currently, you can either (1) find and edit a template manually, or (2) use a service on which you arrange the elements of your website (e.g., text, images, media) in an intuitive way and the service takes care of the HTML and CSS for you. 

Then, you must make the hosting and domain name decisions. Quite simply, your web host is the place where all the files necessary to create your website are stored and your domain name tells web servers where to find the files for your website. A good analogy is that the domain name is the address of your house and the web hosting is the actual house that address points to\footnote{wpbeginner.com,\href{https://www.wpbeginner.com/beginners-guide/whats-the-difference-between-domain-name-and-web-hosting-explained/}{What’s the Difference Between Domain Name and Web Hosting (Explained)}, February 14th, 2019}. If you would like to understand hosting and domain names better, I recommend reading  \href{https://www.wpbeginner.com/beginners-guide/whats-the-difference-between-domain-name-and-web-hosting-explained/}{this article} which explains their differences and interaction.

Another decision to make is what cost you are willing to pay for your website. The cost will depend on the features that you want (on your website and in the creating process), the ease of creation and maintaining your website, and whether you want a customized domain name. It is possible to pay nothing for your website, but this typically requires you to devote more time to the process and get your hands dirty (e.g. choosing a template, understanding HTML and CSS to customize your page, finding a host and domain, and maintaining your page). The costs typically arise if you want (A) a one stop shop (templates, hosting, and domain name) and/or a drag and drop that doesn't require much to any knowledge of HTML/CSS, or (B) a personalized domain name (such as \href{emilymoravec.com}{emilymoravec.com}). 

There are many options for the three main components of your website and choosing can be overwhelming. After I created my own website, I decided to ask the astronomical community to get a sense of what others had done in a survey. The goal of this document is to be a resource for early career scientists looking to make their own personal, professional websites and provide you with a place to start. The layout of this document is as follows. In section \ref{sect:survey}, I describe the survey that I conducted and report the results. In section \ref{sect:options}, I describe various options for the framework to generate website content, hosting, and registering domain names. In section \ref{sect:advice}, I provide advice for content and general tips. Lastly in section \ref{sect:concl}, I summarize and describe my choices.

\section{Survey Responses}\label{sect:survey}
I wanted to know what services and frameworks scientists in the astronomical community had used to create their professional websites. Thus, in August 2020 I created a survey and posted it to the `Astronomers' Facebook page and my LinkedIn, and emailed it directly to a few colleagues. Over the next month, I received a total of 54 responses from 23 (46\%) people in an academic position beyond postdoc (faculty, scientist, etc.), 1 (2\%) individual research fellow, 4 (7\%) people in their third postdoc, 4 (7\%) people in their second postdoc, 16 (30\%) people in their first postdoc, and 6 (11\%) graduate students. 53 of the participants were astronomers and one was a computer scientist. I do not claim that this survey reflects the full distribution of services and frameworks used by the entire scientific or astronomical community. This survey was instead intended to be a vessel to collect ideas for a resource for early career scientists. I also expressed my intent to create a guide for early career scientists with the data from the survey. Below I summarize the questions and responses of interest.

\subsection{Question \#1: At what stage of your career did you create your website?}
The motivation behind this question was to understand when in one's career is it typical to create a personal, scientific website. I will note that it might be the case that a person has created several websites during their career. I did not ask at what stage they first created their website or make it explicit that I was asking about their current one. But it is reasonable to assume that the participants answered about the website they are using currently. 

\begin{figure}
    \centering
	\includegraphics[width=\linewidth, keepaspectratio]{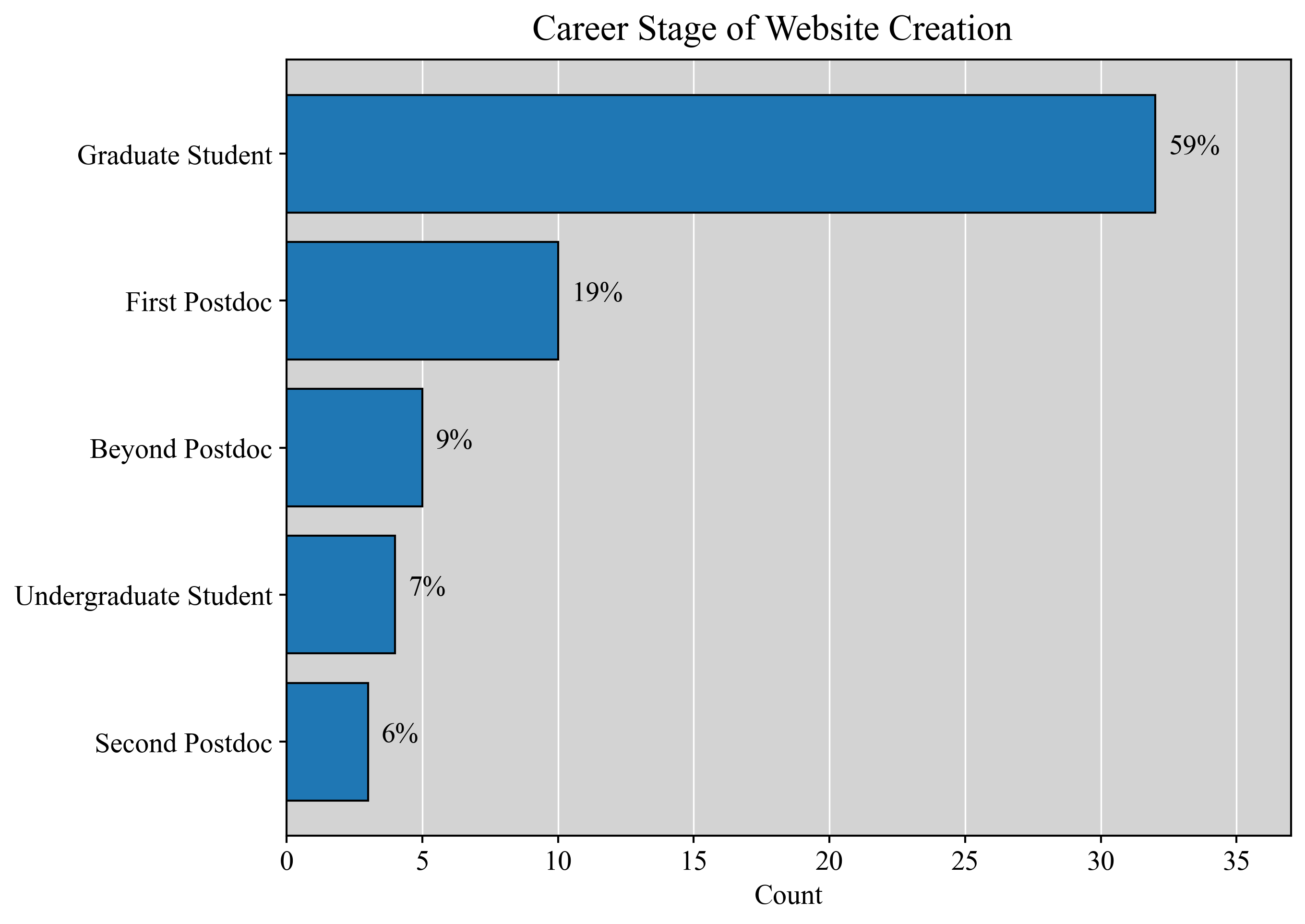}
	\caption{The career stage during which the participants in the survey made their website that they are reporting in the survey.}
	\label{fig:career_stage}
\end{figure}

From Figure \ref{fig:career_stage}, in this sample of scientists it was typical to create a website early on their career when they were a graduate student. I will note that a large majority of the participants in this survey are a postdoc or beyond a postdoc (a more advanced stage of early career and mid career). Since the most popular answer was graduate student, this could represent the typical stage at which scientists in the community create a website. I will also note that the time to complete a PhD varies and thus a senior graduate student and a first or second year postdoc may be the same number of years beyond the beginning of their career. Overall, from this result and personal experience, I recommend to create a website during your time as a graduate student or first postdoc and update it throughout the rest of your career.

\subsection{Questions \#2-4: Hosting, Domain, and Framework}
\begin{figure}
    \centering
	\includegraphics[width=\linewidth, keepaspectratio]{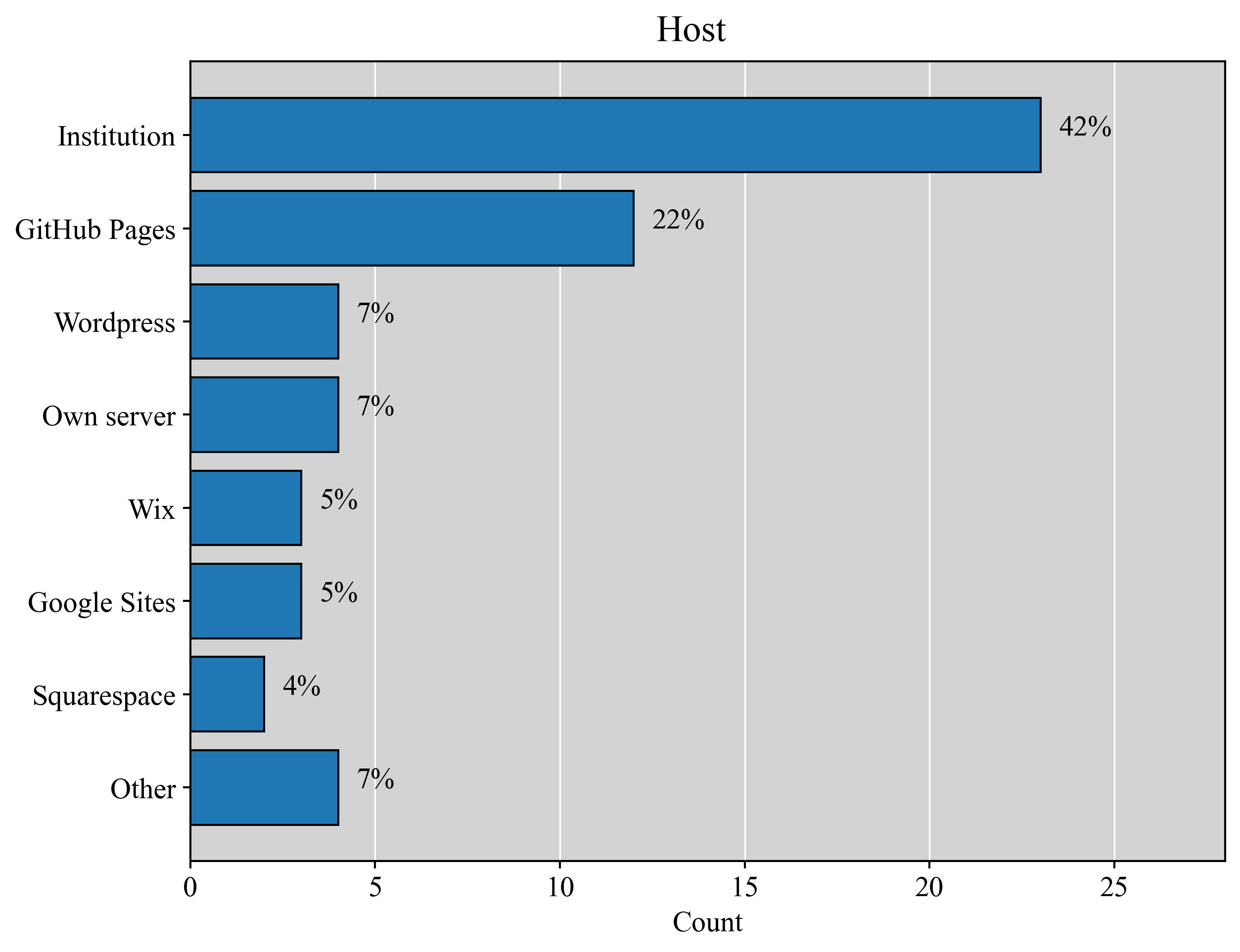}
	\caption{The most popular services the participants in the survey use to host their website. The `Other' category is comprised of all form responses that had only one participant.}
	\label{fig:host}
\end{figure}

\begin{figure}
    \centering
	\includegraphics[width=\linewidth, keepaspectratio]{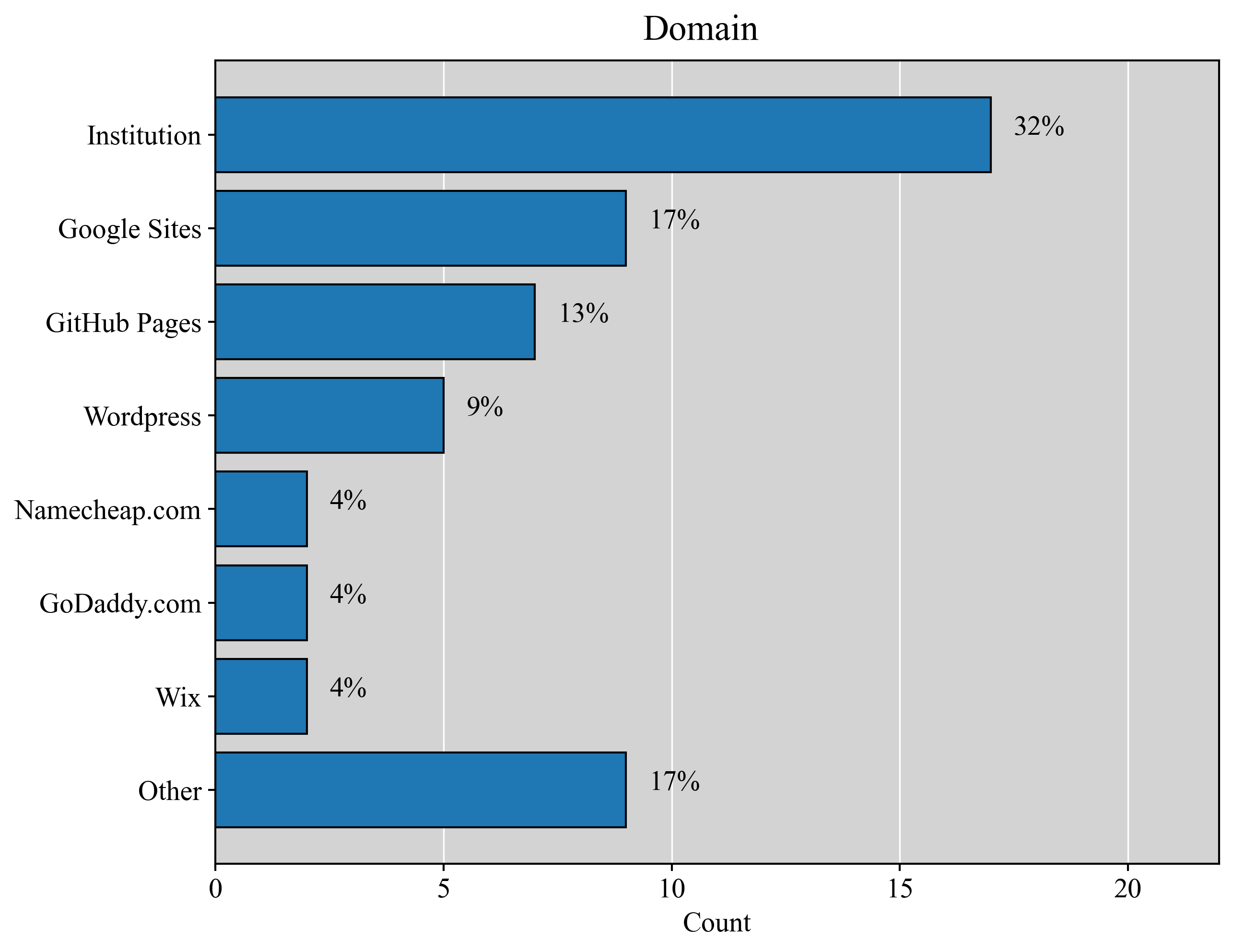}
	\caption{The most popular services from which the participants in the survey obtained their domain name for their website. The `Other' category is comprised of all form responses that had only one participant.}
	\label{fig:domain}
\end{figure}

\begin{figure}
    \centering
	\includegraphics[width=\linewidth, keepaspectratio]{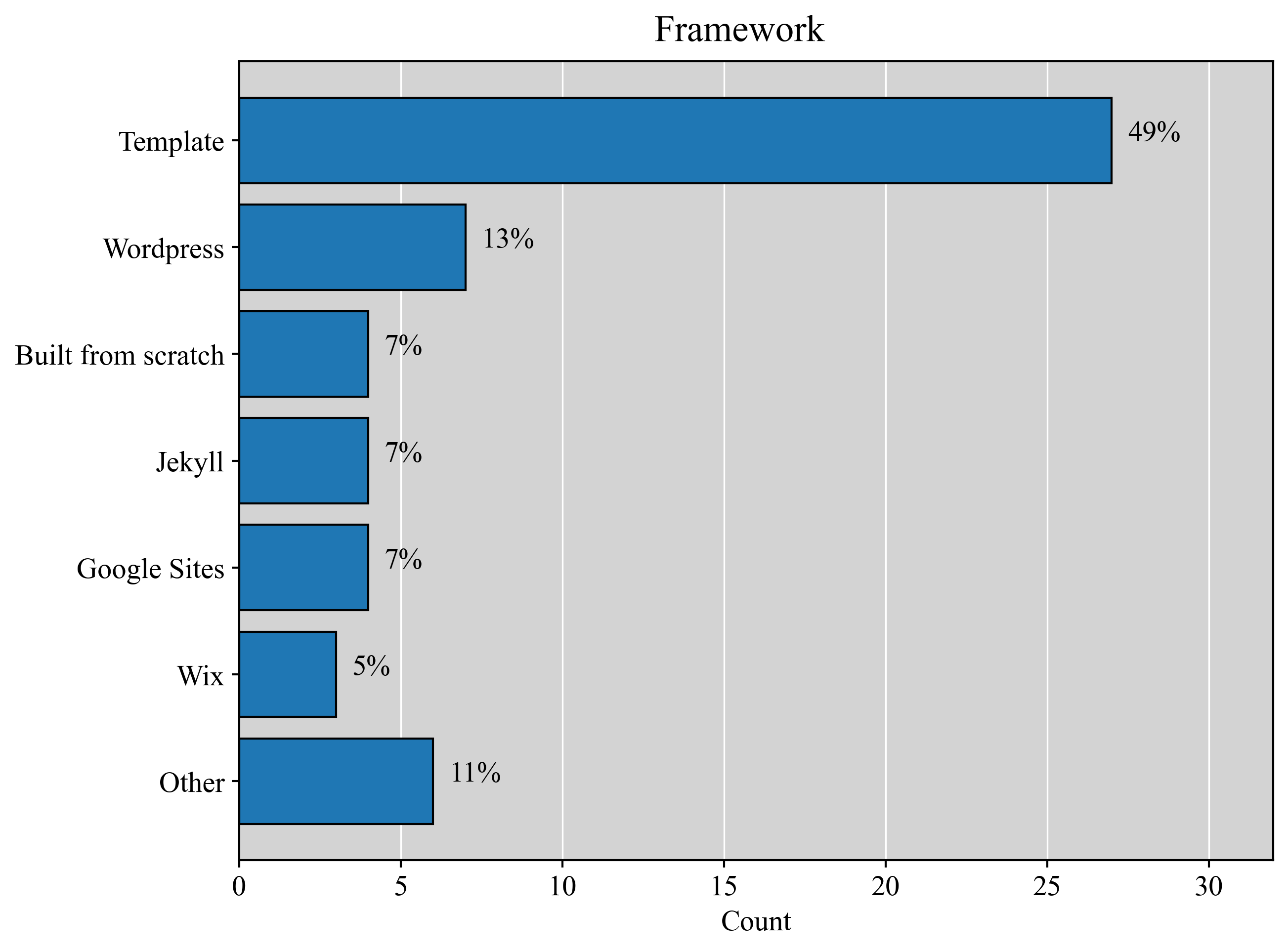}
	\caption{The most popular content generators. `Template' is a HTML (etc.) template. `Built from scratch' is raw HTML (etc.) code with no template. The `Other' category is comprised of all form responses that only two or less participants.}
	\label{fig:framework}
\end{figure}

I asked the following questions in order to get a sense of currently what are the most popular services and frameworks for those in the astronomical community to create a website. 
\begin{itemize}
	\item \textit{Where is your site hosted?} The most popular options for hosting were their institution or GitHub Pages (see Figure \ref{fig:host}). I only display by name those that have two or more responses (see Section \ref{sect:options} for a full list).
	\item \textit{Where did you obtain your domain name?} The most popular options for a domain name were their institution, GitHub Pages, Google Sites, and Wordpress (see Figure \ref{fig:domain}). I will note that $\sim$30\% used a smattering of other services. I only display by name those that have two or more responses (see Section \ref{sect:options} for a full list).
	\item \textit{What web framework did you use to create your site?} The most popular option for generating the HTML code for one's site is to use a template, typically HTML or CSS (see Figure \ref{fig:framework}).I only display by name those that have three or more responses (see Section \ref{sect:options} for a full list).
\end{itemize}
In all three plots, the `Other' category in Figure \ref{fig:host} \& \ref{fig:domain} is composed of all categories that had only 1 response and in Figure \ref{fig:framework} the categories that had two or less responses. Note that there has been some degree of interpretation of the answers to create these plots. All the questions had an option to write in your own answer and I had to interpret the meaning of some of these answers. Additionally, in all three questions one participant put in two answers in fill in your own option. Thus, I added one to each of the respective categories. For hosting and web framework, this meant the total number of responses was 55. For domain name, the total would be 55 but two people could not remember where their domain name came from; thus, the total is 53 in this case. For the domain name, two responses are missing as the participants did not remember where they obtained their domain name for a total of 53. 

Overall, the most popular option is to find a template, edit it, and host it at one's institution or on GitHub Pages. However, I will note that at least four people mentioned in the comments section that if you host at your institution you will have to move your website every time you move institutions. And for early career scientists this is quite often. 


\subsection{Question \#5: How much do you pay per year for your website? (hosting, domain name, web framework, etc.)}
\begin{figure}
    \centering
	\includegraphics[width=\linewidth, keepaspectratio]{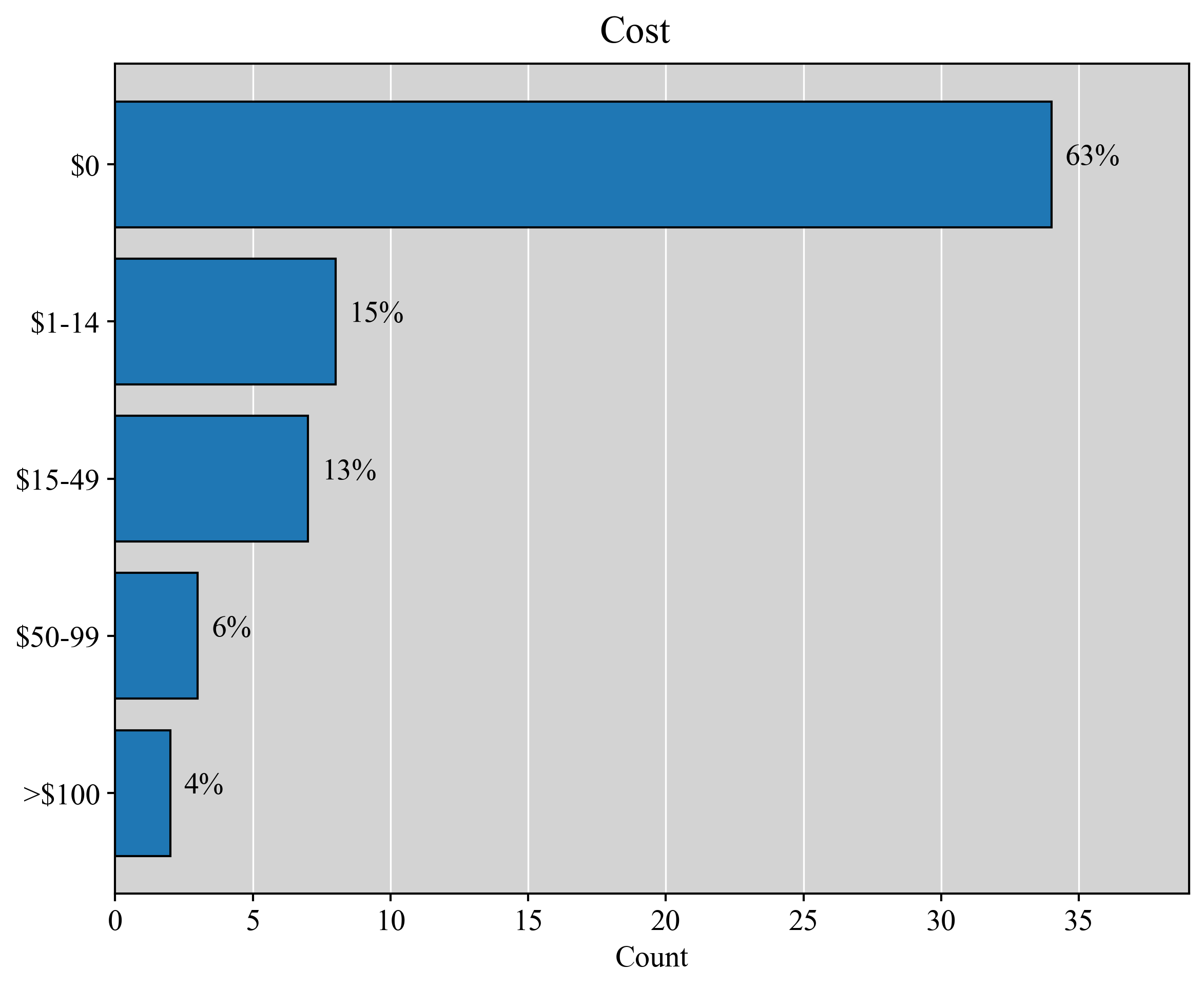}
	\caption{The amount of money that the participants in the survey pay per year for costs associated with their website.}
	\label{fig:cost}
\end{figure}

In Figure \ref{fig:cost}, we see that many people found a way for their websites to cost them nothing and/or maintain one for low cost. This is reflective that many people host their websites on and obtain free subdomain names from services such as their institutions, GitHub Pages, and Google Sites.

\section{Options for Framework, Hosting, and Domain Name}\label{sect:options}
From Figures \ref{fig:host} \& \ref{fig:framework}, the most popular option among the survey participants is to find a template and host it either at their institution, GitHub Pages, or Google Sites. However, I wanted to give an overview of the options for frameworks to generate content, hosting, and domain names that I know of and that the survey made me aware of.

\subsection{"One Stop Shops"}\label{sect:oneStop}
There are several options which provide all three requisites (templates, hosting, and a domain name). Free options include
\begin{itemize}
    \item \href{https://wordpress.com/}{WordPress.com}
    \item \href{https://sites.google.com/new}{Google Sites}
    \item \href{https://www.wix.com/}{Wix}
    \item \href{https://www.weebly.com/?lang=en}{Weebly}
\end{itemize}
Paid options are \href{https://www.squarespace.com/}{SquareSpace} and \href{https://www.hostinger.com/pricing}{Hostinger}. Most of these are  \href{https://en.wikipedia.org/wiki/WYSIWYG}{WYSIWYG} editors which allow the user to drag and drop and do not require much to any knowledge of HTML/CSS to use. That is why they are often preferred by non-programmers, but usually this benefit is at an extra cost or reduced features on a free plan.

These services will often provide a free subdomain name (e.g. username.github.io, name.wordpress.com, sites.google.com/view/yoursite, username.wixsite.com). However, if you want a custom domain (e.g. \href{emilymoravec.com}{emilymoravec.com}), this will cost money.

\subsection{Templates}
Whether you use GitHub Pages, Google Sites, Wix, or your institution to host your site and obtain a subdomain name, you can typically place any HTML or CSS template there. An overwhelming majority of the participants in the survey used a template. Below I list a few sites where you can find free and paid templates:
\begin{itemize}
    \item \href{https://wordpress.com/themes/free}{WordPress}, \href{https://sites.google.com/site/freewebsitetemplategallery/}{Google Sites}, \href{https://www.wix.com/website/templates}{Wix}, \href{https://www.weebly.com/features/website-template?lang=en}{Weebly}, and \href{https://www.squarespace.com/templates}{Squarespace} all have templates available.
    \item Two static site generators that are typically paired with hosting on GitHub Pages are Jekyll (\href{https://jekyll-themes.com/free/}{free} and \href{https://jekyllrb.com/docs/themes/}{all}) and \href{https://themes.gohugo.io/}{Hugo}.
    \item \href{https://startbootstrap.com/templates/}{startbootstrap.com}. Templates that incorporate a responsive design.
    \item \href{https://bootstrapmade.com/}{bootstrapmade.com}. Templates that incorporate a responsive design.
    \item \href{https://www.free-css.com/free-css-templates}{www.free-css.com}
    \item \href{https://html5up.net/}{HTML5 UP!}
\end{itemize}
At many of these sites you can also purchase templates as well.

\subsection{Hosting + Domain Name}\label{sect:hostDomain}
The most popular options for both hosting and domain name from the survey were from `Your Institution' and GitHub Pages. Using your institution for hosting and domain name is free. Ask colleagues or IT staff at your institution if this is an option. If you are interested in using GitHub Pages (also free), see \S\ref{sect:githubPages}. You can also get free hosting and a domain name through those listed in \S\ref{sect:oneStop}. Other options that participants in the survey used to host and register a domain name are \href{https://www.netlify.com/pricing/}{Netlify}(paid), \href{https://www.nearlyfreespeech.net/}{nearlyfreespeach.net}(paid), and their own servers.

\subsection{Domain Name Service (DNS) Providers}\label{sect:domain}
As mentioned before, many of the services where you host your site will provide a free subdomain name (e.g. username.github.io, name.wordpress.com, sites.google.com/view/yoursite, username.wixsite.com). However, if you want a custom domain (e.g. \href{emilymoravec.com}{emilymoravec.com}), this will cost money. You can determine whether your desired domain name is taken using \href{https://www.hover.com/}{hover.com}.

Many of the aforementioned sites that provide templates and hosting services allow you to purchase a custom domain through them (listed in \S\ref{sect:oneStop} and \S\ref{sect:hostDomain}). However, there are other sites through which you can buy and register a domain name that may be cheaper (the `Other' category in Figure \ref{fig:domain}):
\begin{itemize}
    \item \href{https://www.hover.com/}{Hover.com}
    \item \href{https://www.whois.net/}{whois.net}
    \item \href{https://www.domain.com/}{domain.com}
    \item \href{https://www.namecheap.com/}{Namecheap.com}
    \item \href{https://www.ionos.com/domains/domain-names}{IONOS by 1\&1}
    \item GoDaddy.com
\end{itemize}
Many hosting providers will allow you to host with them and specify a customized domain name that you have purcahse. For example, my website is hosted on GitHub Pages and I bought my domain name through Hover.com. The instructions for linking the domain name to hosting provider are available on the hosting providers' help pages.

\subsection{GitHub Pages}\label{sect:githubPages}
I have found that more people are migrating toward using GitHub Pages as it is an easy, free hosting option if you are already familiar with GitHub. However, using GitHub Pages will require that you either (a) upload HTML/CSS template to GitHub or (b) learn Jekyll and use a Jekyll theme. If you don't want username.github.io as your website address, you will have to purchase a custom domain name. And it will require knowing Git to some degree. However, if you are familiar with Git and are willing to spend a bit of time working with some HTML/CSS it is a fairly streamline process. If you decide to do this, below are a few resources:
\begin{itemize}
    \item \href{https://docs.github.com/en/free-pro-team@latest/github/working-with-github-pages/creating-a-github-pages-site}{GitHub Pages How To}
    \item A \href{https://gist.github.com/emoravec/1462145e518a0a2758cc94ada9b987b6}{GitHub gist} that I wrote after creating my site with Jekyll, GitHub Pages, Sublime Text, and a Mac
    \item \href{https://astrosites.github.io/}{Astrosites} which is a GitHub repository maintained by Steven Stetzler and Leah Fulmer
\end{itemize}

\section{Advice and Tips}\label{sect:advice}
In this section I provide general advice concerning website written content and one's web presence gleened from personal experience, survey responses, and discussions with colleagues.
\subsection{Website Written Content}
Below is a list of highly recommended content:
\begin{enumerate}
	\item An introduction page with a photo of yourself, an introduction to your interests, and a short bio.
	\item A shortened CV highlighting the most important items and a downloadable full version.
	\item A list of your publications. Make this prominent. You could include link to your papers on the SAO/NASA Astrophysics Data System (ADS), ORCID, Research Gate, Google Scholar, etc.
	\item A page describing your research in more detail.
	\item Contact information.
	\item Unique content that makes you professionally stand out.
\end{enumerate}

Other content ideas:
\begin{enumerate}
    \item A description of your outreach efforts.
    \item A teaching page. 
    \item A "Science digest" page describing your research in layman's terms for the general public.
    \item A link to your LinkedIn page.
    \item Statistics about your papers, citations, and talks given.
    \item A blog of professional updates.
\end{enumerate}

If you still want more written content ideas, I recommend visiting the websites of colleagues to get ideas. Additionally, it should also be noted that a large fraction of astronomy/astrophysics graduate students may not get a permanent academic job and their websites should also cater to and focus on possibly moving towards non-academic jobs.

A word on linking your publications on your website. For ADS, you can either provide a link to your first author works (e.g. \href{https://ui.adsabs.harvard.edu/search/fq=\%7B!type\%3Daqp\%20v\%3D\%24fq_database\%7D&fq_database=database\%3A\%20astronomy&q=author\%3A(\%22\%5EMoravec\%2C\%20E.\%22)&sort=date\%20desc\%2C\%20bibcode\%20desc\&p_=0}{E. Moravec}) or to all your works. Or you can create an ADS library and provide a link to that. Both have pros and cons. You do not have to remember to update anything when linking to the generic search of your name in ADS. But with the ADS library, you have to remember to keep it updated. But with an ADS librabry can include the most relevant works (no abstracts or proposals if desired). If you create an ADS library make sure you make it public. If you link to your ORCID iD on your website for publications, make sure that you edit the viewing permissions of your works so that public viewers can see it (in the upper right hand corner of each work).

If you are interested in including icons on your website, I recommend \href{https://fontawesome.com/}{fontawesome} for general icons and \href{https://jpswalsh.github.io/academicons/}{Academicons} for academia particular icons. You can also create your own icons, if for example you want to have a customized icon appear in the tab name next to the name of your website. Below are ways to create icons:
\begin{itemize}
    \item Create the icon in PowerPoint, LibreOffice, Keynote/Preview, save as a PNG, then use a web converter to convert the PNG to .ico or .svg formats (e.g. \href{https://convertio.co/png-svg/}{Convertio}). Conversion from PNG to SVG can also be done with GIMP/Photoshop also by opening the file in GIMP and export as SVG.
    \item Create icon in Adobe Illustrator or its open-source equivalent, Inkskape.
    \item One can make arrows and other shapes in pure CSS.
    \item Use unicode symbols (by far the easiest option for a lot of cases).
\end{itemize}.
A quick word on the work required to make your website. Expect to spend an entire week or two creating the initial content and figuring out your hosting and domain work.

\subsection{Browser Development Tools and Responsive Design}
It can sometimes be difficult make certain aspects of your website look exactly as you would like them to. A tool that can help you identify and change a particular element on your website is your \href{https://developer.mozilla.org/en-US/docs/Learn/Common_questions/What_are_browser_developer_tools}{browser's developer tools} (devtools). The editor is particularly useful because it displays the HTML on your page at runtime and the CSS that is applied to each element on the page. Additionally, it allows you to modify the HTML and CSS locally in the browser and view the results of these changes.

You can also use the inspector to test what your website looks like viewed from many different types of devices using the `Toggle device toolbar' (typically in the upper left hand part of the menu bar). This will allow you to determine if you are using responsive design which simply means that your websites resize properly when viewed on different types of devices. I would recommend picking a template that has a responsive design (e.g., \href{https://startbootstrap.com/templates/}{bootstrap templates}).

\subsection{General Web Presence and Website Advice}
In this section, I provide advice on maintaining a professional web presence and various aspects of website creation.
\begin{enumerate}
	\item \textbf{Web presence:} In general, it is a good idea to clean up your web presence. A reason to create a website is to control what comes up when your name is searched. What do you want a future employer to see when they search your name? Apart from having a professional website, it is also important to have an awareness about what is visible about you online. It may not matter how impressive your professional website is, if in the meantime the second Google search about you is unprofessional content and photos. I recommend that you Google yourself in an incognito tab and see what comes up. Have your friends Google you and report back. Then remove unprofessional content. There are procedures for all search engines for removing unwanted contact (e.g., \href{https://support.google.com/websearch/troubleshooter/3111061?hl=en}{removing personal information from Google}).
	\item \textbf{LinkedIn:} I recommend getting a LinkedIn page early on in your career. I learned a lot about efficient use of LinkedIn from talks by \href{http://www.alainalevine.com/}{Alaina Levine}. I recommend \href{https://www.youtube.com/watch?v=BuJOT86ccic\&list=PLFhVT3VzlwKrbuWhFsVaYUcnb1ZDeW91r\&index=5}{this video} where she spoke to members of the American Astronomical Society about using LinkedIn. One thing that she recommends is \href{https://www.linkedin.com/help/linkedin/answer/87/customize-your-public-profile-url?lang=en}{customizing your LinkedIn url}.
	\item \textbf{How do you ensure that when someone googles your name, your website comes up?} This is a field in and of itself called search engine optimization (SEO). But I have a few recommendations. First, make sure keywords that people will put in the search bar are on your website. For me, this would be "Emily Moravec radio astronomy." Second, if you choose to use a template, choose one that includes metadata tags for your pages. In Jekyll themes this is the \texttt{title} and \texttt{description} variables. Third, once you make your website, send it to colleagues, friends, and family to get some hits on your website (plus you can get their feedback). Sometimes it simply takes time for enough traffic to go to your site for it to appear on google search. It can take several weeks to appear among the top recommendations on Google. Fourth, have a link to your website everywhere you can to accrue more hits (e.g. LinkedIn, email signature, talks, your institution's website, etc.). If you want more information about SEO see \href{https://www.seomechanic.com/why-is-my-website-not-showing-in-google-search-results/}{this article}.
	\item \textbf{Website:} Register your site with google analytics and get a tracking ID number. You will be able to analyze your site traffic this way.
	\item \textbf{Website:} Right after your website goes live request that colleagues and friends visit your website, test the links, and give general feedback on it.
	\item \textbf{Website:} Keep the content on your website simple, clear, and up to date.
	\item \textbf{Website:} Be aware that getting things to look \textit{exactly} how you want will take time and effort and can be a huge time sink. Be flexible.
\end{enumerate}
In the survey I created I asked, "Do you have an advice for an early career scientist creating their website? Or something you wish you had known before you started the website creation process?" All of the answers are listed in the Appendix. But in general, the responses revolved around putting up a \textit{simple} website that conveys your information clearly and keep it updated. It is interesting that a large percentage of the participants use their institution for hosting and a domain name, however four participants point out that even though it is free to use your institution to host and get a domain name, you will have to deal with moving it from institution to institution. They recommend to host it somewhere other than your institution then you won't have to move it when starting a new position.

\section{Conclusion} \label{sect:concl}
There are many options for creating a website to promote your work which can be overwhelming. Through a survey of 54 participants and my own experience, I provide a broad overview of the options currently available to early career scientists who desire to create a professional website to promote their work. The conclusions that I take away from this project are:
\begin{enumerate}
    \item Create a website early in your career (recommended is graduate school). Having a website is better than having none. What do potential employers and colleagues find when they search your name?
    \item Keep your website simple, clear, and updated.
    \item Include an introduction to your interests, keywords about you and your work, some version of your CV, publications list, and a summary of your work.
    \item Be prepared to invest time into making your website (1-2 weeks). 
    \item Many in the astronomical community have found a way to create a website that costs $\lesssim$\$15 per year.
\end{enumerate}

Personally, I chose to use a Jekyll template, host on GitHub Pages, and purchase a personalized domain (\href{emilymoravec.com}{emilymoravec.com}). I wanted to have a website that I did not have to move with me from institution to institution. Additionally, I am familiar with git and I was willing to take some time to learn how Jekyll worked, thus GitHub Pages was the best long-term solution for me. If you would like to do this too, I have written \href{https://gist.github.com/emoravec/1462145e518a0a2758cc94ada9b987b6}{a git gist} about this or you can follow the tutorial provided by \href{https://astrosites.github.io/index.html}{AstroSites}. I also purchased a personalized domain name from \href{Hover.com}{Hover.com} for \$15 per year. 

One of the intents of this document was to layout many of the popular options for creating one's own professional website. However, the reader could very well still be feeling overwhelmed with options. How do I choose? My best advice is to look over the options presented in this document, and then ask your colleagues how they created their websites and what they recommend. Then decide for yourself which option best suits you. 

\acknowledgements
E.M. would like to thank all of those who filled out the survey. E.M. would also like to thank Thomas Chamberlin, Peter Boorman, Stevie Carnell, and Abhijeet Borkar for discussions and contributions to this document.

\appendix
The final question of the survey was, "Do you have any advice for an early career scientist creating their website? Or something you wish you had known before you started the website creation process?" Below are the responses from the participants that responded to this question:
\begin{itemize}
    \item Don't pay for services like Squarespace if you have the time to figure out free services like GitHub.
    \item Pick something that is easy to maintain and does not a lot of updating. Less is more. Use automatically generated publication lists such as google scholar.
    \item \href{https://astrosites.github.io/index.html}{AstroSites} (a github repo maintained by Steven Stetzler and Leah Fulmer) is a fantastic step by step resource to put together a website quickly!
    \item Make it easily accessible and readable. Someone shouldn't need to click through multiple pages/links to get to your papers, or CV, or institution, etc. (Think: Will it load on conference wifi?)
    \item Simple can be effective! A site can grow later, but get basic information and a link to an ADS page of your work out there on the web to start.
    \item It is much better to have your own site not connected to your institution since you will likely be changing institutions a few times. Also, keep your site updated regularly!
    \item Put something together that's bare bones and has your CV, interests, and background information. Then you can slowly build something more custom as time goes on. I had "build website" on my to do list for at least a year before I finally pulled the trigger, and it only took a few hours to actually do something basic.
    \item Spend time to make something you’re proud of. It is worth it in the long run.
    \item Keep it professional, there are other platforms to host your favorite pictures, links, hobbies, etc.
    \item Up to date, clear, CV with academic history (for talk intros), email.
    \item The best moment to create your website is at graduate school.
    \item Host on github, edit an HTML5 template, pull to your institutions website if you can.
    \item Make it easy for you to make small updates frequently, rather than one big update every year. Keep professional and personal websites/social media accounts separate. I want to learn about your research/teaching/service, not your parties, politics, etc.
    \item Just make it basic and informative. People looking to hire you are mostly looking for research highlights, activities (outreach, public lectures), and publication information.
    \item Keep it simple, don’t overload with information.
    \item Create yours as early as possible, and shamelessly advertise it.
    \item Don't be too cute or edgy with your design - you'll get sick of it. Simple, classic, and easy to update as styles change is best.
    \item A bit more information--I've had my own website since 2007 but I didn't start paying for my own domain until my second postdoc. I had always hosted it through my academic institution. That is the advice I would give--your academic institution website won't be permanent.
    \item Make a website on a free service and just see what you like and don't like, then start fresh on the real one having already thought about things (and delete the old one from the internet!)
    \item Find a template you like and modify it. Have it hosted somewhere other than your institution, so you don't have to copy it over when you move institutions!
    \item Creating a website is itself super easy. But be mindful that it costs money (sometimes monthly) to maintain the website on a server and to keep ownership of the domain name. Also be sure to list things like teaching experience and outreach.
    \item It took me much longer than expected to get things to look the way I wanted, I think it would have been really helpful to draft a layout before starting digging into the HTML and CSS codes. Also the W3C school website really helped me a lot in terms of creating specific tools and appearance. I was using a template from html5up, it was a bit hard to get started since I knew nothing about HTML, but it ended up being a great learning experience.
    \item It's not a bad idea to pay for a domain name and learn some basic HTML templates. It goes a long way relative to a google site/wix site.
    \item Find something that's easy to maintain and update. Beyond that it all depends on individual needs.
    \item Have a website. Any website is better than no website. Also this is just personal preference, but I actually like simpler websites. Usually, all I need is a photo ("is this the person I met at the thing"), three sentences about what you do ("yes this is the person"), your email, and links to your CV and ads page. I really don't care to scroll for five minutes through your entire life story. (but clearly other people have the opposite opinion, so do what works for you).
    \item Make it early, make it simple. No watermarks or ads. Don't host on institutional sites until you're in a permanent position: use http redirects or links to point to your personal off-institutional domain. Don't let it get out of date, even between job seasons - it is not uncommon for people with grants to recruit in off-seasons. Not having a website can be a warning that you're not interested or available in academic jobs; you don't need a good website, though, you just need a site with CV plus contact info.
\end{itemize}




\end{document}